# Solidification of the Glass-Forming Al$_{86}$Ni$_2$Co$_6$Gd$_6$ Melt under High Pressure

S.G. Menshikova[a,*], N.M. Chtchelkatchev[b], V.V. Brazhkin[b]

[a]Udmurt Federal Research Center, Ural Branch, Russian Academy of Sciences, Izhevsk, 426034, Russia
[b]Institute of High Pressure Physics, Russian Academy of Sciences, Moscow, Troitsk, 108840, Russia

**ABSTRACT.** High pressures allow the synthesis of new metastable compounds that remain intact for a sufficiently long time at normal conditions. Until now, it has not been fully understood how pressure, glass-forming ability and solidification of liquids are interconnected. We have investigated the structure of the glass-forming eutectic alloy Al$_{86}$Ni$_2$Co$_6$Gd$_6$ obtained by rapid cooling from the melt having a temperature of 1800 K under a pressure of 10 GPa. X-ray diffraction analysis and electron microscopy show that the samples are homogeneous and dense. The structure is finely dispersed. New stable crystalline phases with cubic (cP4/2) and tetragonal (tI26/1) structures are formed in the alloy. The studies have shown that the average microhardness of the samples obtained at 10 GPa is almost 2 times higher than that of the original sample at atmospheric pressure and is about 1700 MPa. To understand the results, we used ab initio molecular dynamics and studied how the melt changes with pressure. It is shown that at a temperature of 1800 K, high pressure increases the concentration of icosahedral clusters in the melt so that at 10 GPa atoms inside the icosahedra form a percolation cluster, while at atmospheric pressure they do not. Thus, the glass-forming ability of a melt increases at high pressure strongly influencing solidification processes.

## 1. Introduction

High pressure induces important changes during the solidification of alloys [1] such as the formation of metastable phases and changes in strength and hardness [2, 3]. Another factor that strongly influences crystallization process is glass-forming ability. Recently, it has been found that the nucleation of crystals in a supercooled melt usually reaches the maximum rate at the glass transition temperature and it extremely slows down during deep cooling in the glassy state [4]. We have performed experimental and theoretical studies to understand how solidification occurs under high pressure and at the high glass-forming ability of a melt.

Alloys of the Al-TM-REM type (TM - transition metal; REM - rare-earth metal) containing 80-90 at.% Al, in particular, alloys of the Al-(Ni,Co)-REM eutectic type, are quite easily amorphized by the melt-spinning method at ultra-rapid quenching from the liquid phase [5]. However, the small thickness of amorphous ribbons (~20-40 μm) limits their practical application. In contrast to the so-called bulk amorphous alloys based, for example, on zirconium, iron, and gold [6, 7], for amorphous ribbons from the chosen alloys of the AL-TM-REM type there is the necessity to increase the ribbon critical thickness, and this is a problem to date. One of the criteria for the bulk amorphization of alloys is the presence of phases with a complex crystal structure such as «tau» - phases for metal-metalloid systems or Laves phases for metal-metal systems. The formation of these phases from a melt is difficult due to the high energy of their nucleation; therefore, when the cooling rate is increased, it is easily suppressed, which contributes to easy amorphization. To date, there are no exact empirical rules and theoretical foundations for choosing specific alloy compositions in the Al-TM-REM system that will have the maximum amorphizing ability. One of the promising approaches in this direction is the search for such compositions using high (several GPa) and ultrahigh (hundreds of GPa) pressures. Such pressures have a significant effect on the thermodynamics and kinetics of the melt solidification, which can additionally increase the tendency to the bulk amorphization of the Al-TM-REM alloys. It is well known that the glass-forming properties of melts are largely determined by the kinetics of collective processes of structure formation. However, similar processes observed in the eutectic glass-forming melts are still poorly understood and do not have a satisfactory theoretical description. Thus, the progress in understanding the processes of the structure formation occurring in glass-forming melts of the Al-TM-REM type in the eutectic region, as well as in studying the effect of pressure and cooling rate on the processes of their solidification, is of high scientific importance.

---

* Corresponding author.
*E-mail address*: svetlmensh@mail.ru (S.G. Menshikova)

The rapid development of computer technology makes it possible to conduct theoretical studies of realistic systems using various methods, in particular, within the framework of the ab initio approach (AIMD) [8]. The ab initio approach is the solution of a problem from first fundamental principles without involving additional empirical assumptions. A direct solution of the equations of quantum mechanics is implied. In this case, some assumptions and simplifications are made. Such simplifications enable to study systems with a relatively large number of atoms or atoms with a large number of electrons. Recently, the methods of ab initio calculations based on the density functional method have become more and more widespread in condensed matter physics [9]. The advantage of the ab initio calculations is the exact description of the atomic interaction taking into account quantum effects. The main drawback is the impossibility of calculating microscopic systems with a large number of particles, for example, atoms (rarely more than 1000) in a reasonable time.

As applied to the selected type of alloys, the effect of high pressure on the local structure of a melt, as well as on the formation of structures during the melt solidification, has not been practically studied in the literature. The purpose of the present work is to study the possibility of the formation of new phases in the $Al_{86}Ni_2Co_6Gd_6$ alloy during the rapid solidification of its high-temperature melt under high pressure, as well as to study the structural features of the melt.

## 2. Materials and methods

The samples under a high pressure of 10 GPa were obtained in a toroid-type high-pressure chamber [10]. Alget stone and BN ampoule were used as a pressure-transmitting medium. The sample was heated and melted by passing an alternating current through it. The temperature value was calculated based on the thyristor readings and according to the current power. The melt was cooled at a rate of 1000 deg/s. The melt temperature before quenching was 1800 K. The scheme of the experiment is as follows: setting the pressure → pulsed heating → holding at the set pressure and temperature → cooling to room temperature without depressurizing → decreasing the high pressure to atmospheric. The phase composition of the samples was determined by X-ray diffraction analysis on a Dron-6 setup using Co-$K_\alpha$ radiation. The spectra taking modes were selected so that the sufficient accuracy of the determination of the angular position of the diffraction peaks and integrated intensity would be achieved. For determining the phase composition, lattice parameters, phase space group, the X-ray diffraction patterns were processed using the PHAN program from the MIS&A package. The profiles of the X-ray diffraction patterns of the studied samples were processed and analyzed with the use of the above programs. The system Quattro S - Scanning Electron Microscope (SEM) with a standard detector DBS (directional backscatter detector) ABS/CBS was used to determine the chemical and elemental composition, morphology and size of the structural components of the alloy. The error in determining the percentage of elements in the samples is no more than 5%. Durametric measurements (Vickers hardness, $H_v$) were performed on a PMT-3M microhardness tester. The load applied to the indenter was 20 g and the indentation time was 10 s. The $H_v$ values were averaged over 20 measurements.

The calculations were performed by the density functional method based on the VASP package (The Vienna Ab initio Simulation Package) [11]. We considered elementary cells consisting of 512 atoms with periodic boundary conditions at the Gamma point. The cutoff energy of the plane wave basis was chosen to be equal to 500 eV. Due to the low concentrations of Ni, Co, and Gd, the simulation of the melt by the quantum molecular dynamics (QMD) method was performed using 10 independent QMD trajectories with different initial random arrangement of atoms. Random initial configurations were created based on the interaction potential of hard spheres between atoms and classical (LAMMPS) molecular dynamics simulation [12]. Then the most disordered configurations (special quasirandom structures) were selected using USPEX package [13]. The equilibrium configuration was achieved by the QMD simulation of the system in the NPT ensemble for at least 10 ps with a step of 1 fs. The calculation approach is described in detail in [14, 15]. To investigate liquid transformation at deep overcooling we built the machine learning interaction potential (MLIP) using quantum molecular dynamics data as the training dataset [16]. We used DEEPMD package [17] as the core to build MLIP. DEEPMD is designed to minimize the effort required to build deep learning-based model of interatomic potential energy and force field and to perform classical molecular dynamics (CMD) [18]. We chose se??? a descriptors with cut-off 7 (A) to process the training data. The resulting accuracy of MLIP was 0.02eV/A for forces and 1 meV for energy compared to ab initio calculations. Classical molecular dynamics with MLIP has accuracy comparable to quantum molecular dynamics; CMD is also more than three orders of magnitude faster than QMD. The most important is that CMD can deal with a large enough number of particles and propagate them within micro second time periods, rather than 1-10 nano seconds. We selected LAMMPS [19] as the CMD engine. For CMD simulation of the liquid alloy we took the cell with 13824 atoms in the periodic boundary conditions and the time step equal to 2 fs. The alloy was simulated under cooling using Nose-Hoover NPT thermostat from temperature T=2000 (K) down to 100 (K) for 10 million MD steps.

## 3. Results and discussion
### 3.1. Experiment

High pressure affects the liquidus temperature of alloys. For the selected alloy, the liquidus temperature was determined at high pressures of 5 and 10 GPa (Fig. 1). Thus, as the pressure increases to 10 GPa, the temperature shifts by ~ 200 degrees relative to the equilibrium melting temperature of the alloy. For aluminum, the dependence of the liquidus temperature of the alloy on high pressure is more pronounced: the displacement is ~500 degrees (Fig. 1, [20]). For the studies under high pressure, the melt temperatures were taken paying attention to the above displacement.

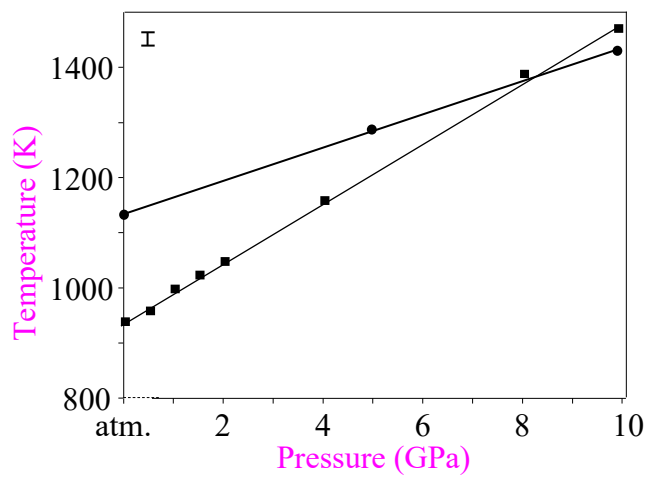

**Fig. 1.** The melting points of Al (■, [20]) and the $Al_{86}Ni_2Co_6Gd_6$ alloy (●, our data)

Table 1 presents the synthesis conditions and the phase composition of the studied samples determined by X-ray diffraction analysis. In the samples obtained by quenching from the melt at atmospheric pressure only equilibrium phases are present (see Table 1). In the alloy obtained under 10 GPa, two new phases are formed, namely, the Al₃Gd phase (the Al₃U type) containing Ni and Co and having a primitive-cube structure (cP4/2) with a lattice parameter a=0.4285±0.002 nm, and the Al₈Co₄Gd phase (the Al₈Cr₄Gd type) having a tetragonal structure (tI26/1) with parameters a=0.8906±0.0003 nm and c=0.5150±0.0003 nm. Figure 2 shows the X-ray diffraction patterns of the sample with the new detected phases. Previously, the samples from the studied alloy were obtained under pressures of 3, 5, and 7 GPa [21]. The new phases start to form in the alloy already at 7 GPa.

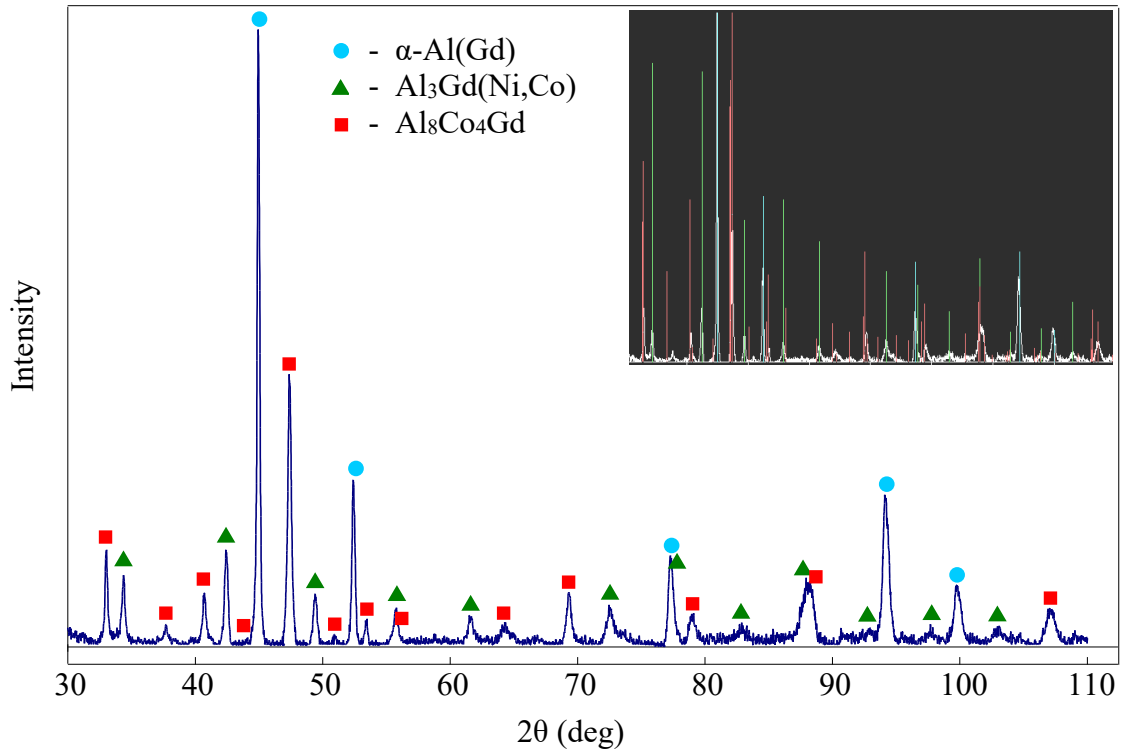

**Fig. 2.** X-ray diffraction patterns of the alloy (10 GPa, 1800 K)

Table 1
The preparation conditions and the phase composition of the samples

| № | Condition of preparation | Phase composition |
|---|---|---|
| 1 | atm. press. (1800 K) | α-Al(Gd) (cub., cF4/1), Al₉Co₃Gd₂ (orth., oC56/7), Al₃Gd (hex., hP8/3), Al₃Ni (orth., oP16/2) |
| 2 | 10 GPa (1800 K) | α-Al(Gd) (cub., cF4/1), Al₃Gd(Ni/Co) (the Al₃U type) (cub., cP4/2), Al₈Co₄Gd ( the Al₈Cr₄Gd type) (tetr., tI26/1) |

Figure 3 shows the microstructure of both samples considered in this work. As can be seen from Fig. 3, the structure of the sample obtained under 10 GPa is finely dispersed and uniform. In addition, no pores or shrinkage cavities are found. In the composition of the α-Al solid solution, the content of Gd is almost 2 times higher than that in the sample obtained at atmospheric pressure. Figure 4 shows the concentration maps of the distribution of elements in the samples.

The studies have shown that the average microhardness of the sample obtained at 10 GPa is almost 2 times higher than that of the original sample at atmospheric pressure and is about 1700 MPa. The increase in strength is due to solid solution and dispersion strengthening.

## 3.2. Ab initio Calculations

To obtain a deeper insight into the studied system we performed ab initio calculations. As a result of quantum molecular dynamics simulation of the melt at 1800 K and at pressures 0 and 10 GPa, the radial distribution functions (RDF) were obtained and studied. Figure 5 shows the total RDF. Figure 6 presents partial RDFs of the melt at 0 and 10 GPa (1800 K) for aluminum, cobalt, nickel and gadolinium. The increase of the peaks and their shift indicate an increase in the degree of the local ordering in the melt at high pressure. A decrease in the average value of the interatomic distance is observed. As can be seen from Fig. 6, high pressure mainly affects the local environment of aluminum and gadolinium in the melt. To analyze the short-range order in the arrangement of atoms, a method known as "polyhedral template matching" [22] was used with a RMSD cutoff 0.2, which is standard for a liquid. The analysis of the molecular dynamics trajectories by this approach shows that at zero pressure, icosahedra in a small concentration of about 0.5% are present in the melt. With an increase in pressure to 10 GPa, their concentration increases to at least 4%. In this calculation of the icosahedra concentration only atoms having an icosahedral nearest surrounding are taken into account, i.e. at the center of an icosahedron. Figure 7 shows all atoms in the $Al_{86}Ni_2Co_6Gd_6$ melt contained in icosahedral clusters (not only at the center of the icosahedron, but also in its shell) at 0 GPa (a) and 10 GPa (b). For convenience, the atoms are surrounded by a surface (shown in green). The surface around the atoms is built according to a standard algorithm. The algorithm is contained in the calculation program OVITO (The Open Visualization Tool)[†] [23]. The main criterion used at the building of such surface is as follows: it is necessary that the curvature radii at each point of the surface are slightly larger than the characteristic sizes of the atoms. In this case the surface is sufficiently smooth on atomic scales. If we take into account the remaining atoms in the first coordination sphere (there are 12 atoms in this case; see Fig. 7) and assume that the icosahedra are isolated, then it turns out that at 10 GPa about 48% of the atoms are included in the icosahedra (such a number of atoms already form a percolation network), and at 1 GPa only 12%. Thus, at 10 GPa the atoms form a percolation cluster, while at atmospheric pressure they do not. For both of the above pressures these calculations are performed at 1800K. If at P=0 we cool the system by 200 K, from 1800 K to 1600 K, after the decrease of pressure from 10 to 0 GPa, the concentration of icosahedra will be below 2%. Thus, pressure is essential here. The network of icosahedra in a metallic liquid is usually a mark of a glass forming ability. Consequently, as the pressure increases, the glass-forming ability of the melt also increases.

It is known that similarly to the phase second-order transitions, during glass formation, an abrupt change in heat capacity and thermal expansion coefficient occurs. In this case, a kink is observed on the temperature dependence of the volume. Figure 8 shows the evaluation of the dependence of glass transition temperature on pressure based on the temperature dependence of volume per one atom. In Fig. 8, the bending point is the glass transition temperature. The glass transition temperature determined by computer modeling (about 600 K under normal pressure and 820 K at pressure of 10 GPa) makes up about 0.5 of the temperature of melting (see Fig. 8), which is in good agreement with the literature data for similar metal systems. As is known, glass transition temperature weakly (logarithmically) depends on cooling rate. At the experimental rates of cooling $10^6 - 10^7$ K/s, which are 5 orders of magnitude lower than the effective cooling rate at the computer modeling (in the present paper, $7.5*10^{11}$ K/s), the glass transition temperature will be several tens of degrees lower. Both the considerable growth of the glass transition temperature with increasing pressure and a little growth of the relation $T_g/T_m$, namely, from 0.52 to 0.57, should be noted; this may indicate an increase in the tendency toward glass formation.

---

[†] The Open Visualization Tool (OVITO) is a new 3D visualization software designed for post-processing atomistic data obtained from molecular dynamics.

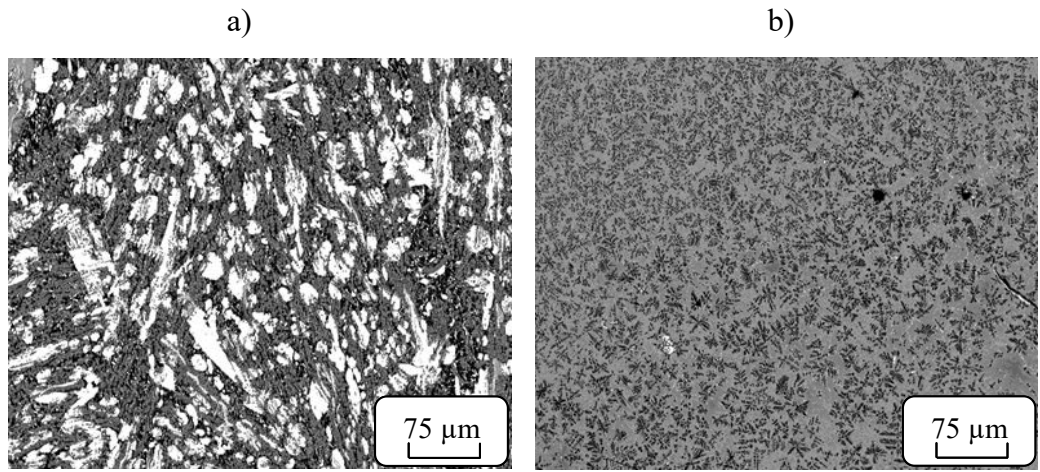

**Fig. 3.** Microstructure of the samples synthesized at normal conditions (a) and at high pressure of 10 GPa (b) at temperature 1800 K and cooling rate 1000 deg/s.

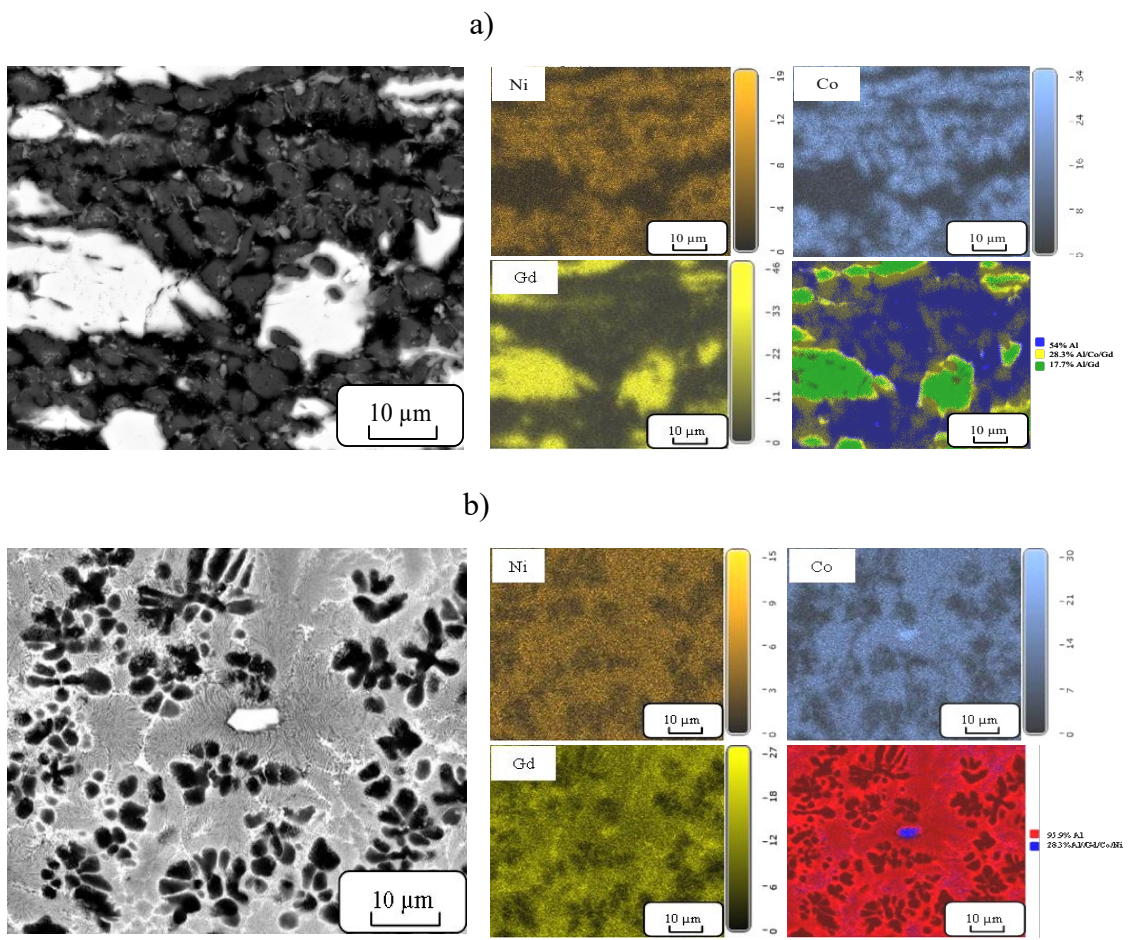

**Fig. 4.** Concentration maps of the distribution of elements in the samples at atmospheric pressure (a) and 10 GPa (b) (1800 K, 1000 deg/s).

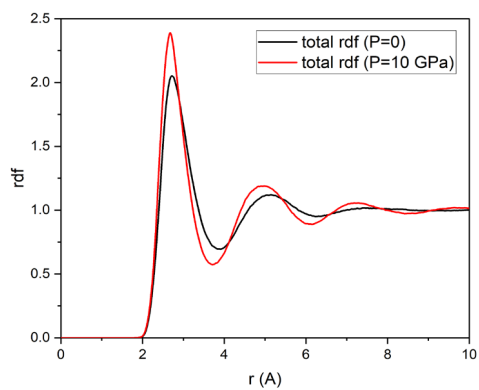

**Fig. 5.** Total RDF of the melt (1800 K)

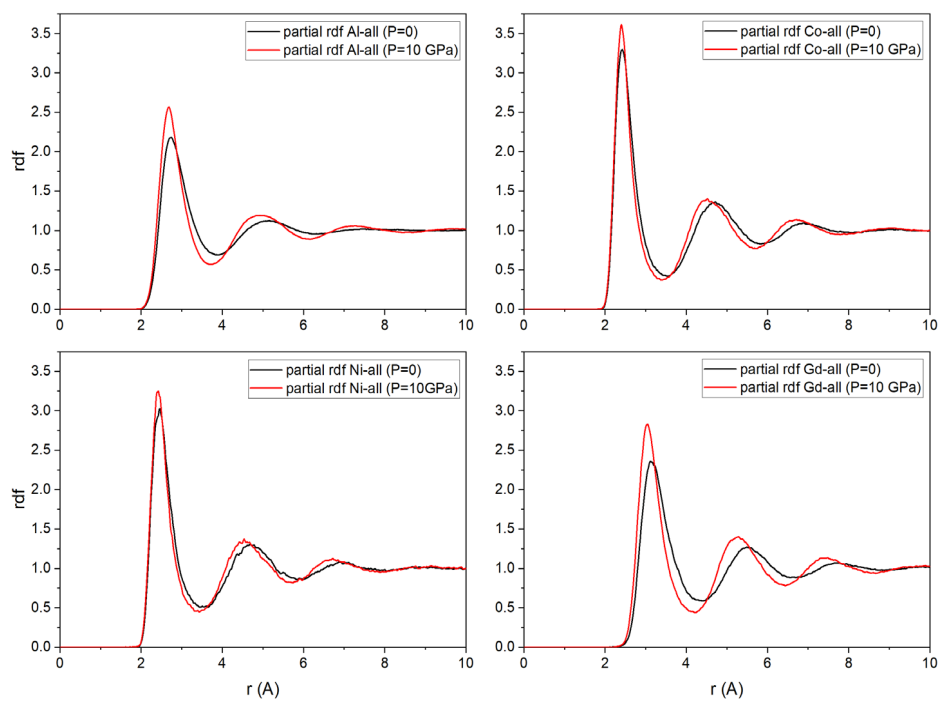

**Fig. 6.** Partial RDFs of the melt (1800 K)

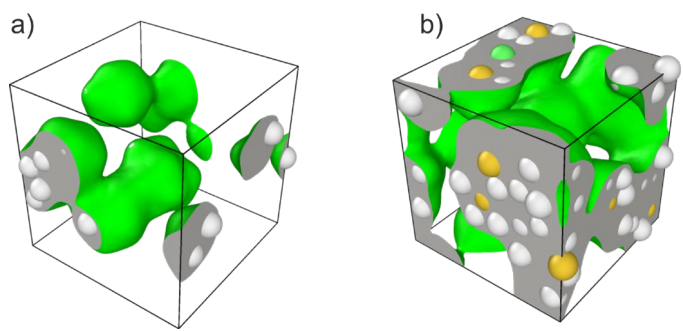

**Fig. 7.** Atoms within icosahedral clusters in the $Al_{86}Ni_2Co_6Gd_6$ melt are shown (other atoms are removed) at pressures: 0 GPa (a) and 10 GPa (b) and temperature 1800 K.

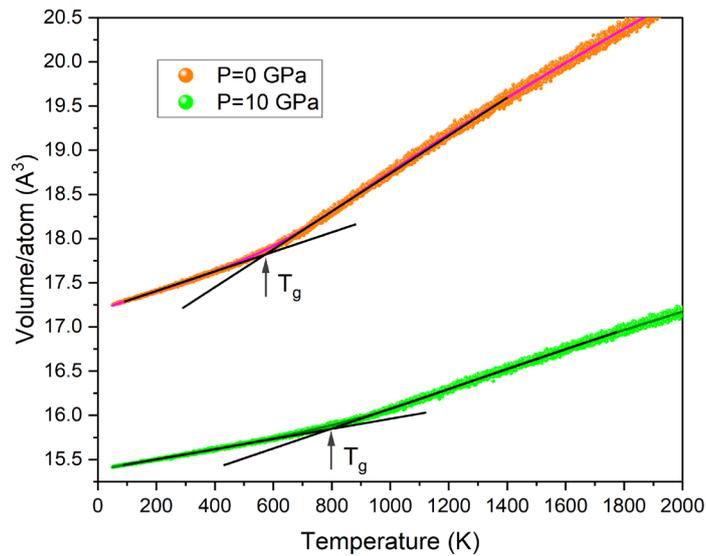

The interrelation between the external conditions and the state of a melt influence the solidification of the melt; this interrelation should be taken into consideration during the development of technological processes for obtaining materials with specified properties. High pressures affect both the kinetics and thermodynamics of the melt solidification. On the one hand, pressure increases the viscosity of the metal melt and, on the other hand, it modifies the structure of the short-range order in the direction of denser packings, including non-crystalline ones. In addition to the general tendency towards grain refinement, pressure also promotes the crystallization of new metastable phases which are usually denser. The crystallization of the melt occurs due to the formation of crystalline nuclei growing into crystals. Spontaneously appearing centers of crystallization (nuclei) lead to a decrease in the volume free energy and increase in the interphase energy due to the interface. The formation of crystal nuclei in a liquid requires certain conditions. The appearance of micro-volumes is necessary where the mutual arrangements of atoms would correspond to the crystal lattice of hard alloy. As is shown in the calculations given in our paper, high pressure primarily greatly influences the local environment of Gd and Al in the melt. This is thermodynamically efficient for the formation of primary crystals of a new metastable phase $Al_3Gd$ in a solid sample; the new phase has a more symmetric cubic lattice in comparison with the known equilibrium phase $Al_3Gd$. Further, in the conditions of compression an anomalously supersaturated solid solution Al(Cu) and the second new phase are formed. The phases obtained under high pressure are metastable (the state of the obtained phases is non-equilibrium); however, they are sufficiently stable. We show the presence of a percolation cluster under high pressure in the studied metal melt, which usually provides an increase in the glass-forming ability of melts; however, we have found no amorphous phases in the considered melt. Apparently, for the glass formation in melt much higher pressures are necessary at the same (considered) cooling rate.

The results obtained show the fundamental possibility of using the method of melt solidification under high pressure to change the properties of aluminum alloys used in industry (without changing their chemical composition by modifying the structure and changing the composition of the structural components of the sample). They make a significant contribution to understanding the processes of structure formation occurring in glass-forming melts of the Al-TM-REM type in the eutectic region, as well as to studying the effect of high pressure on the processes of their solidification.

## 4. Conclusions

The combination of high rates of a high-temperature melt solidification and a high pressure of 10 GPa leads to the formation of new crystalline phases in the $Al_{86}Ni_2Co_6Gd_6$ alloy: $Al_3Gd$ (of the $Al_3U$ type) containing Co and Ni and having a primitive-cube structure (cP4/2) and $Al_8Co_4Gd$ (of the $Al_8Cr_4Gd$ type) having a tetragonal structure (tI26/1); in addition, the α-Al solid solution is anomalously supersaturated with gadolinium. The structure of the sample is fine-crystalline and of high density. The average microhardness is high due to solid solution and dispersion strengthening.

High pressure mainly affects the local environment of gadolinium and aluminum in the melt. The study of the short-range order shows that there are a certain number of icosahedra in the melt. At zero pressure, the number of icosahedra are about 0.5%, while with the increase in pressure to 10 GPa, it increases to 4%. These features lead to an increase in the glass forming ability of the melt under increasing pressure. In addition, new phases are formed in the $Al_{86}Ni_2Co_6Gd_6$ alloy. A deeper theoretical study of the system based on the machine learning for constructing interatomic potential will allow to consider much longer spans of time in numerical simulation and to investigate solidification processes directly. This will be discussed in our further papers.

## Declaration of Competing Interest

The authors declare that they have no known competing financial interests or personal relationships that could have appeared to influence the work reported in this paper.

## Acknowledgments

The work was supported by the Russian Science Foundation (Project No. 22-22-00674). The electron microscopic studies were performed using equipment of the core shared research facilities of the "Center of physical and physical-chemical methods of analysis, investigation of properties and characteristics of surface, nanostructures, materials and items" of the UdmFRC of the UB RAS, Izhevsk. The numerical calculations were performed using computing resources of the federal collective usage center Complex for Simulation and Data Processing for Mega-science Facilities at NRC "Kurchatov Institute" (http://ckp.nrcki.ru/), supercomputers at Joint Supercomputer Center of RAS (JSCC RAS), and the "Govorun" supercomputer of the Multifunctional Information and Computing Complex, LIT JINR (Dubna).